# LANNDD-- A Massive Liquid Argon Detector for Proton Decay, Supernova and Solar Neutrino Studies, and a Neutrino Factory Detector[*]

David B. Cline, UCLA
University of California Los Angeles,
Department of Physics and Astronomy, Box 951447
Los Angeles, California 90095-1547 USA

Franco Sergiampietri, UCLA/PISA
Istituto Nazionale di Fisica Nucleare
Via Livornese, 1291
56010 San Piero a Grado (PI) Italy

John G. Learned, University of Hawaii, Manoa
High Energy Physics Group
Department of Physics & Astronomy
329 Watanabe Hall
2505 Correa Road
Honolulu, Hawaii 96822

Kirk McDonald, Princeton
Experimental High-Energy Physics
Department of Physics
PO Box 708
Princeton, New Jersey 08544

**Abstract.** We describe a possible Liquid Argon Neutrino and Nuclear Decay Detector (LANNDD) that consists of a 70kt magnetized liquid argon tracking detector. The detector is being designed for the Carlsbad Underground Laboratory. The major scientific goals are:

1) Search for $p \to k^+ + \overline{\nu_\mu}$ to $10^{35}$ years lifetime
2) Detection of large numbers of solar neutrino events and supernova events
3) Study of atmospheric neutrinos
4) Use as a Far Detector for a Neutrino Factory in the USA, Japan or Europe

## 1. Introduction/Motivation

One option for a next generation nucleon decay search instrument is a fine-grained detector, which can resolve kaons as well as background from cosmic ray neutrinos that are below the threshold for water Cerenkov detectors such as Super-Kamiokande (SK). Such a detector can make progress beyond the *few* $\times 10^{33}$ yr limits from SK for SUSY favored modes because the reach improves linearly with the time and not as the square root of exposure as in SK. This is because the background is low and the detection efficiency high. It will be possible to discover nucleon decay up to about ~$10^{35}$ yr lifetime/branching ratio with an instrument of ~70 kT mass in liquid argon after a few years of exposure.

A second major goal for such an instrument, as demonstrated in a spectacular example of synergy in the last two generations of underground detectors, is the study of neutrino interactions and oscillations. Such a detector can make neutrino oscillation studies using the cosmic ray neutrinos alone (being able to resolve muon neutrino regeneration, detect tau's and tighten measurements of $\Delta m^2$ search for other mixing than $\nu_\mu \to \nu_\tau$). But coupled with a neutrino factory, this detector, outfitted with a large magnet, offers the advantage of being able to discriminate the sign not only of muon events, but of electron events as well. Given the bubble-chamber-like ability to resolve reaction product trajectories, including energy/momentum measurement and excellent particle identification up to a few GeV, this instrument will permit the study of the neutrino MNS matrix in a manner which is without peer.

---

[*] For the Neutrino Factory meeting at KEK - NUFACT 01

One may question whether such a marvelous instrument is affordable, by which we mean buildable at a cost comparable or less than the neutrino source cost. It is indicated by simple scaling from existing experience with ICARUS, that such an instrument will cost out in the class of a large collider detector instrument and represents a straightforward extrapolation of existing technology.

As expected for such a large, isotropically sensitive, general-purpose detector, there are *many* ancillary physics goals which can be pursued. This device would allow exploration of subjects ranging from the temporal variation of the solar neutrino flux (above a threshold of perhaps 10 MeV), to searches for neutrinos from individual or the sum of all supernovae and other cataclysmic events (e.g. GRBs), to cosmic ray research (composition, where the WIPP depth is advantageous), dark matter searches (via annihilation neutrinos), searches for cosmic exotic particles (quark nuggets, glueballs, monopoles, free quarks), and point source neutrino astronomy. In all these instances, we can go beyond SK by virtue of lower energy threshold, better energy loss rate resolution, momentum, angle, sign and event topology resolution.

We note that in the "Connecting Quarks with the Cosmos: Eleven Science Questions for the New Century" report (now in draft, 1/9/01, http://www.nas.edu/bpa/reports/cpu/index.html) for the National Academy, that such an instrument addresses eight of the eleven questions at least indirectly and of those eight, two explicitly (nucleon decay and neutrino mass).

## 2. The Carlsbad Site for LANNDD

In Figure 1 we show the possible location of LANNDD at the Carlsbad Underground National Laboratory site (CUNL). Note that the ease of construction and the exhaust pipe are key motivations for this site. Safety would be accomplished by walling off the detector from the rest of the lab. Excavation is relatively inexpensive at this site due to the salt structure.

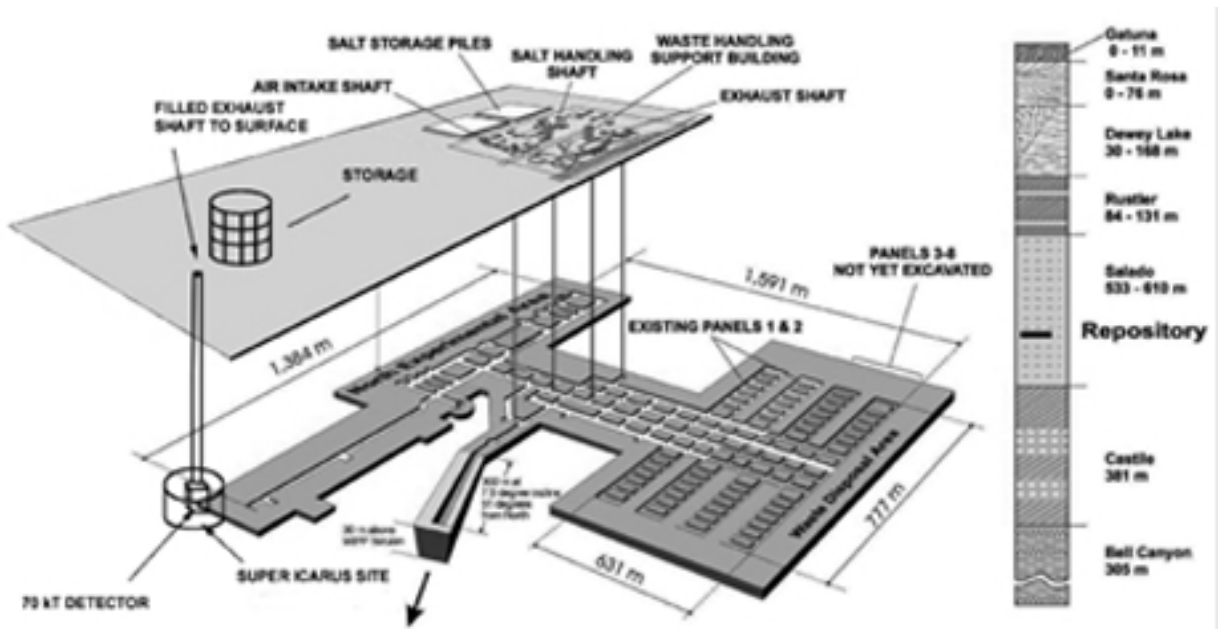

Figure 1. Liquid Argon Neutrino and Nuclear Decay Detector at the CUNL site.

### 3. Some Scientific Goals of LANNDD

Much of the scientific studies to be done with LANNDD follow the success of the ICARUS detector program (Ref. 1, 2, 3). The main exception is for the use of the detector at a neutrino factory where it will be essential to measure the energy and charge of the $\mu^{\pm}$ products of the neutrino interaction. We will soon propose an R&D program to study the effects of the magnetic field possibilities for LANNDD.

a) *Search for proton decay to $10^{35}$ years*

The detection of $p \to k^{+} + \overline{\nu_{\mu}}$ would seem to be the key channel for any SUSY-GUT model. This channel is very clean in liquid argon due to the measurement of the range and detection of the decay products. We expect very small background at even $10^{35}$ nucleon years exposure for this mode (refer to ICARUS studies).

b) *Solar neutrinos and supernova neutrinos studies*

The major solar neutrino process detected in liquid argon is:

$$\nu_e + {}^{40}Ar \to {}^{40}Ar^{*} + e^{-}$$

with *Ar\** de-excitation giving photons with subsequent Compton events. The same process is useful for supernova $\nu_e$ detection – the expected rate for the solar neutrinos is ~123,000 per year. For a supernova in the center of the galaxy with full mixing there would be ~3000 events – no other detectors would have this many clean $\nu_e$ events.

c) *Atmospheric neutrino studies*

By the time LANNDD is constructed it is not clear which atmospheric neutrino process will remain to be studied. However this detector will have excellent muon, hadron and electron identification as well as the sign of the $\mu^{\pm}$ charge. This would be unique in atmospheric neutrino studies.

The rate of atmospheric neutrinos in LANNDD will be (50KT fiducial volume):

  Electron Neutrino Events  4800/year

  Muon Neutrino Events  3900-2800/year (depending on the neutrino mixing)

There would also be about 5000 neutrino current events/year. We would expect about 25 detected $\nu_{\tau}$ events per year that all would go upward in the detector.

d) *Use of LANNDD in a neutrino factory*

Because of the large mass and nearly isotropic event response, LANNDD could observe neutrinos from any of the possible neutrino factories: BNL, FNAL, CERN or JHF in Japan. There are two approximate distances $(2-3) \times 10^{3}$ km and $(7-8)\,9 \times 10^{3}$ km for these neutrino factories. We assume the more distant neutrino factories operate at 50 GeV $\mu^{\pm}$ energy. For a neutrino factory that produces $10-20\mu^{\pm}$ per year at FNAL/BNL and expect ~50,000 per year of right sign μ (i.e. $\mu^+ \to e^+ + \nu_e + \overline{\nu_\mu}$: the $\overline{\nu_\mu}$ gave $\mu^+$ as right sign muons.

The number of wrong sign muons will depend on the mixing angle $\theta_{13}$ (the wrong sign muon is $\mu^+ \to e^+ + \nu_e + \overline{\nu_\mu}$; $\nu_e \to \nu_\mu$, $\nu_\mu \to \mu^-$: the $\mu^-$ is the wrong sign muon)---there could be as many as 5000 wrong sign events/year.

For the farther distances (CERN in Japan) these numbers would be about the same due to the higher energy $\mu^{\pm}$ (50 Gev) with the rate increasing like the $E_\mu$ to the 3$^{rd}$ power.

The LANNDD detector could be useful for the search for CP violation from any neutrino factory location. This will depend on the value of the mixing angle $\theta_{13}$ and the magnitude of the CP violation.

For the longer distance experiments (CERN/Japan) there could be an important MSW effect that is important to study in order to evaluate the significance of the CP violation search. It is possible that the electric charge of the $e^{\pm}$ from the reaction $\overline{\nu_\mu} \to \overline{\nu_e}$ could be determined as it was in heavy liquid bubble chambers by following the shower particles---this is currently under study.

In Figure 2 we show the schemata of neutrino factory beams to the CUNL site for detection by LANNDD---we consider this a universal neutrino factory detector.

## 4. The Detector and the magnitude of the CP violation.

The aim is to build a 70 kT active volume liquid argon TPC immersed in magnetic field. The geometric shape of the detector is mainly decided by the minimization of the surface-to-volume ratio S/V, directly connected to the heat input and to the argon contamination. Spherical (diameter=D), cubic (side=D) or cylindrical (diameter=height=D) shapes have all the minimum S/V (=6/D). As compromise between easy construction and mechanical stability, the cylindrical shape has been preferred. Adding to the S/V criterion the need of minimizing the number of readout wires (=electronic channels) and of maximizing the fiducial-to-active volume ratio, a single module configuration appears definitely advantageous with respect to multi-module array configuration (see Table 1). The more difficult mechanical design for the single volume configuration appears fully justified by the larger fiducial volume, the lower number of channels, the lower heat input and contamination and then lower construction and operating costs.

Table 1. *Fiducial volume, number of channels and heat input (calculated with 1 W/m$^2$) for different detector configurations.*

|  | Single Module | 8 Modules | 64 Modules |
|---|---|---|---|
| Active volume, m$^3$ | 50000 | 50000 | 50000 |
| Fiducial volume, m$^3$ | 41351 | 33559 | 21037 |
| Number of channels | 164261 | 337787 | 724077 |
| Heat Input, W | 9104 | 18209 | 36417 |

The internal structure of the detector is mainly relied to the maximum usable drift distance. This parameter depends on the acceptable attenuation and space diffusion of the drifting charges. Acceptable working conditions are obtained with an electric field of 0.5 kV/cm, a drifting electron lifetime of 5÷10 ms and a maximum drift of 5m. The detector appears then as sliced into 8 drift volumes (Figures 2, 3), 5 m thick, each confined between a cathode plane and a wire chamber. Each wire chamber is made of two readout planes (*u, v*) with wires oriented at +45° and −45° with respect to the horizontal plane. A 5 mm wire pitch gives a sufficiently detailed imaging of ionizing tracks in the drift volumes.

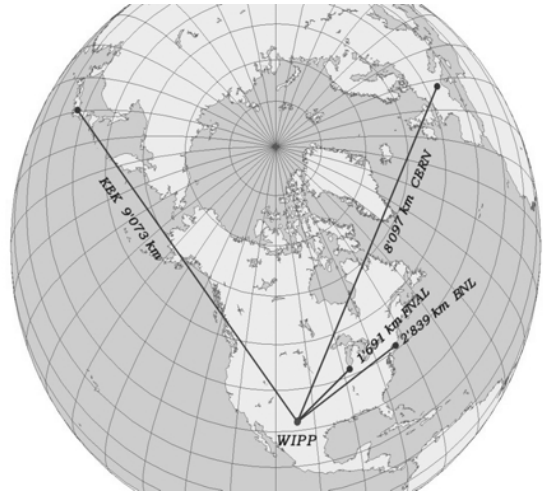

**Figure 2. Schematic of the possible neutrino factory beams to the CUNL site.**

| LBL Beam | Distance | $\theta_H$ | $\theta_V$ |
|---|---|---|---|
| FNAL | 1 691 km | 47.9° | 7.8° |
| BNL | 2 839 km | 62.5° | 12.9° |
| CERN | 8 097 km | 41.6° | 39.5° |
| KEK | 9 073 km | - 44.7° | 45.4° |

$\theta_H$: angle, in the WIPP site horizontal plane, respect to the NORD
$\theta_V$: angle, at the WIPP site, respect to the horizontal plane

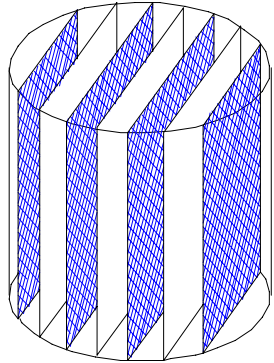

Figure 3. *Schematic layout of chamber (hatched regions) and cathodes planes (white regions).*

The magnetic field is vertically oriented and is obtained with a solenoid around the cryostat containing the liquid argon. With such an orientation the maximum bending for a charged particle is obtained in a horizontal plane and appears in the imaging as an arc in each of the planes ($u, t$) and ($v, t$).

The detector is foreseen as located underground (Figure 4), at a depth of 655m (2150 ft) in a housing equipped with an emergency liquid argon pool and with argon vapor exhaust ducts. Forced fresh air inlet, liquid/vapor nitrogen in/out ducts, assembling hall with crane and elevator complete the basic organization of the underground cave. The magnetic field is vertically oriented and is obtained with a solenoid around the cryostat containing the liquid argon. With such an orientation the maximum bending for a charged particle is obtained in a horizontal plane and appears in the imaging as an arc in each of the planes ($u,t$) and ($v,t$).

As the difficulties of this project rely mainly on its engineering and safety aspects, a realistic mechanical design with costs and construction time estimates is matter of a dedicated feasibility study to be approved and properly funded.

For the full project definition a preliminary activity is required to study *a)* the imaging in a magnetized liquid argon TPC, *b)* the operation in conditions of high hydrostatic pressure, *c)* drift path of $\geq$ 5m.

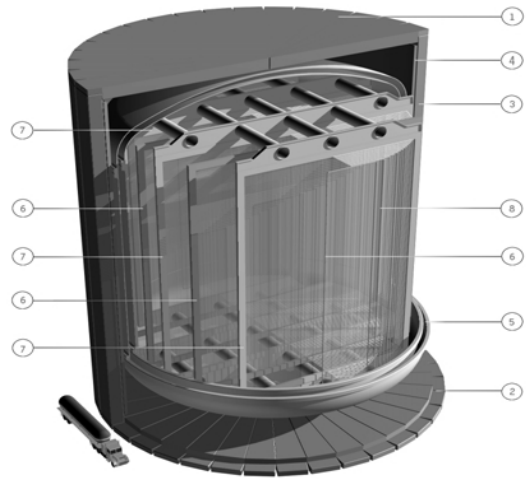

Figure 4. *Artistic view of the preliminary sketch for the LANNDD detector: 1) Top end cap iron yoke; 2) Bottom end cap iron yoke; 3) Barrel iron return yoke; 4) Coil; 5) Cryostat; 6) Cathodes; 7) Wire chamber frames; 8) Field shaping electrodes.*

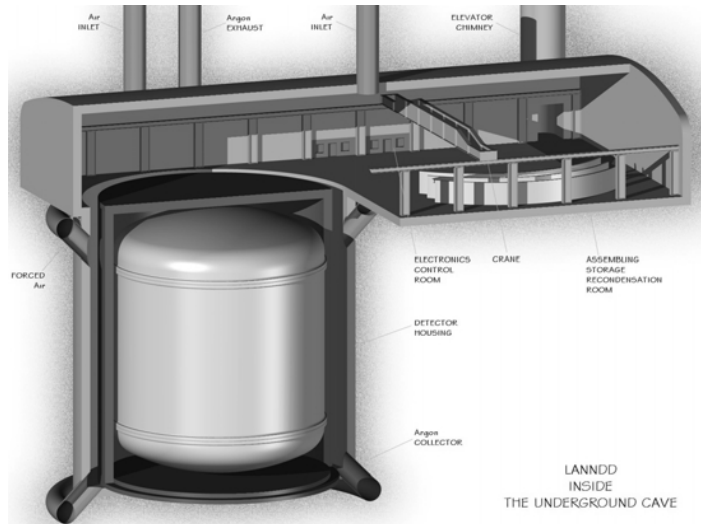

Figure 5. *The LANNDD inside the underground cav*e

**References**

[1] ICARUS Collaboration, *"ICARUS-II. A Second-Generation Proton Decay Experiment and Neutrino Observatory at the Gran Sasso Laboratory"* , Proposal Vol. I & II, LNGS-94/99, 1994.

[2] ICARUS Collaboration, *"A first 600 ton ICARUS Detector Installed at the Gran Sasso Laboratory"* , Addendum to proposal, LNGS-95/10 (1995).

[3] F. Arneodo et al. [ICARUS and NOE Collaboration], *"ICANOE: Imagine and calorimetric neutrino oscillation experiment"* , LNGS-P21/99, INFN/AE-99-17, CERN/SPSC 99-25, SPSC/p314; see also A. Rubbia [ICARUS collaboration], hep-ex/0001052. Updated information can be found at http://pcnometh/cern/ch.